\documentclass[aps,preprint,showpacs,showkeys]{revtex4}%
\usepackage[dvips]{color}
\usepackage{amsfonts}
\usepackage{amsmath}
\usepackage{amssymb}
\usepackage{graphicx}%
\providecommand{\U}[1]{\protect\rule{.1in}{.1in}}

\begin{document}
\preprint{ }
\title[quantum information storage]{Raman scheme for adjustable bandwidth quantum memory}
\author{J.-L. Le Gou\"{e}t}
\affiliation{Laboratoire Aim\'{e} Cotton, CNRS UPR3321, Univ. Paris Sud, b\^atiment 505,
campus universitaire, 91405 Orsay, France}
\author{P. R. Berman}
\affiliation{Michigan Center for Theoretical Physics, and Physics Department, University of
Michigan, Ann Arbor, Michigan \ \ 48109-1040}

\begin{abstract}
We propose a scenario of quantum memory for light based on Raman scattering.
The storage medium is a vapor and the different spectral components of the
incoming signal are stored in different atomic velocity classes. One uses
appropriate pulses to reverse the resulting Doppler phase shift and to
regenerate the signal, without distortion, in the backward direction. The
different stages of the protocol are detailed and the recovery efficiency is
calculated in the semi-classical picture. Since the memory bandwidth is
determined by the Raman transition Doppler width, it can be adjusted by
changing the angle of the signal and control beams. The optical depth also
depends on the beam angle. As a consequence the available optical depth can be
optimized, depending on the needed bandwidth. The predicted recovery
efficiency is close to 100$\%$ for large optical depth.

\end{abstract}
\volumeyear{year}
\volumenumber{number}
\issuenumber{number}
\eid{identifier}
\date{\today}
\startpage{1}
\endpage{ }
\maketitle





\section{Introduction}

The storage of quantum information in an atomic ensemble has received a great
deal of attention over the past ten years or so. Protocols based on
electromagnetic induced transparency (EIT) have been investigated by many
groups both theoretically \cite{fleisch,fleisch2} and experimentally, leading
to storage and retrieval demonstration of both discrete \cite{chan,eisa} and
continuous \cite{appel,honda,cvik} quantum variables. Although successful, the
EIT storage scheme suffers from time-bandwidth product limitations. In
EIT-based protocols, the reduction of group velocity is used to spatially
confine the input signal within the boundaries of the storage medium.
Simultaneously, the signal spectrum must not exceed the bandwidth of the
transparency window associated with EIT. Since the group velocity is inversely
proportional to the width of the transparency window, it follows that, the
larger the storage bandwidth, the shorter the temporal profile the memory can
accommodate. This protocol has been demonstrated using level schemes in which
inhomogeneous broadening does not play a significant role. Extension to
inhomogeneously broadened systems does not improve the time-bandwidth product capabilities.

On the other hand, inhomogeneous broadening can be of critical importance in
other protocols for storing quantum information. For example, Doppler
broadening determines the storage bandwidth when a signal pulse is totally
absorbed in an optically dense medium \cite{mois}. While the temporal
components of the input signal are distributed in atomic state coherence along
the axial direction in EIT, the spectral components of the input signal are
spread over the inhomogeneous frequency distribution of the atoms in the
absorption protocol. Just as in EIT, information is stored in a long lifetime
Raman coherence in the absorption protocol, but the maximum duration of the
input signal is independent of the storage bandwidth, being ultimately limited
only by the inverse homogeneous line width. The resulting time-bandwidth
product capacity, given by the ratio of the inhomogeneous and homogeneous
widths, is reminiscent of the photon-echo based storage techniques that were
developed in the past \cite{lin}. However, unlike some of those classical
light storage schemes, the proposal in Ref. \cite{mois}, that we shall refer
to as the MK protocol, is restricted to systems where the inhomogeneous
broadening is provided by the Doppler effect.

A variant of the MK protocol has been proposed for solids in which the
inhomogeneous broadening is linked to stochastic variations in atomic
transition frequency that depend on an absorbing center's position in a host
medium \cite{nils, kraus}. Although there are several proof of principle
experiments of this absorption-type protocol \cite{hetet1,hetet2,alex}, all
such experiments have involved classical input fields. It should be stressed
that the large time-bandwidth product capacity is lost when the Doppler shift
is replaced by a more stochastic source of inhomogeneous broadening. In order
to keep control of the inhomogeneous phase shift, one selects a narrow
spectral group of atoms at the beginning. An external field is used to spread
this initial ensemble over the desired bandwidth. Hence the bandwidth is
increased at the expense of the available optical density, i.e. at the expense
of the capacity to trap the optically carried information within the material.

A key feature in the MK scenario is that the totally absorbed signal is
restored without amplification. This contrasts with previous photon echo
investigations where large retrieval efficiency results from strong
amplification in an inverted medium \cite{aza,corn,tsa,wan1,wan2}. Therefore,
unlike these earlier works, the MK protocol is free from the noise associated
with spontaneous and stimulated emission. This is closely related to the fact
that the control fields do not interact with highly populated states. As a
consequence few atoms are promoted to the upper electronic level. Another
consequence is that, as the control fields interact with quasi-empty states,
they are neither attenuated nor distorted as they travel through the active medium.

Quite surprisingly, the original MK scheme has not been demonstrated
experimentally. A possible issue in atomic vapors is the short lifetime of the
active optical transition upper level. Indeed, to combine a large optical
density with the absence of collisions one has to work on strong lines with
short upper level lifetime. As a consequence, some operations have to be
carried out on a nanosecond time scale. Specifically, one needs nanosecond
$\pi$-pulses to convert optical dipoles into Raman coherences and conversely.
In addition, the input signal duration is limited to a few nanoseconds.

In this paper, we propose a Raman variant of the MK scheme that circumvents
these limitations. Direct excitation of ground state coherence avoids the
introduction of rapidly decaying quantities, yet retains the other advantages
of the MK protocol. Quantum memories based on Raman scattering have been
proposed in the past \cite{kozhe,nunn,hetet2}. However, previous proposals did
not fully examine the dynamics of the system, focusing on steady-state
conditions, \cite{kozhe} or they considered a situation in which the temporal
signal profile is mapped into a spatial distribution of the atomic ground
state coherence \cite{nunn,hetet2}. In our scenario, the spectro-temporal
features of the signal are stored in the \textit{spectral distribution} of the
ground state coherence. The experimental investigation presented in ref.
\cite{hetet2} is actually very close to our situation, but the authors resort
to a reversible magnetic field gradient to cover the signal spectrum and to
reverse the atomic phase, which ultimately leads to a spatial mapping of the
signal. In our case we use optical pulses to reverse the atomic phase.

Our objective in this paper is to describe the underlying physics of the Raman
protocol, a goal that can be achieved within the confines of a theory in which
all radiation fields are treated classically. The paper is arranged as
follows: after presenting a picture of the overall process in section II, we
develop a theory for each step of the protocol in sections III-V. In section
VI we discuss the range of applicability of the storage method.

\section{Outline of the storage and retrieval procedure}%

\begin{figure}
[ptb]
\begin{center}
\includegraphics[height=4.0568in,width=4.5031in]
{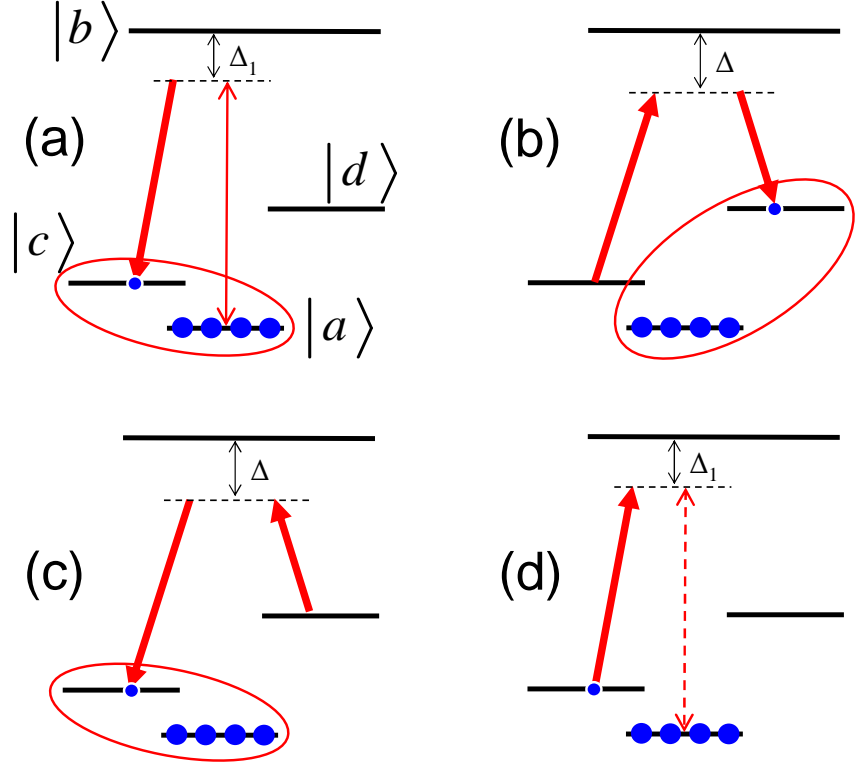}
\caption{(Color online) Level scheme and protocol steps. (a) the weak input
signal, combined with a strong control field, drives the Raman transition
$a-c$. (b) the Doppler phase build-up is stopped by conversion of the
coherence $\rho_{ac}$ into $\rho_{ad}$. This is accomplished by $\pi$-pulse
excitation of the Raman transition $c-d$. (c) the coherence $\rho_{ac}$ is
recovered with the help of a second Raman $\pi$-pulse. (d) a backward
propagating control field creates a coherence $\rho_{ab}$ which allows for the
restoration of the signal pulse, propagating in the backward direction.  }%
\label{fig1a}%
\end{center}
\end{figure}

The protocol consists of the four steps shown schematically in Fig.
\ref{fig1a}. We consider an ensemble of four-level atoms. The four levels can
be magnetic state sublevels of the same or different ground state hyperfine
levels. The atoms are prepared initially in state $\left\vert {a}\right\rangle
.$

\subsubsection{Stage 1}

In the first stage, a quasi-monochromatic control field and the input pulse
drive Raman transitions between levels $a$ and $c$. Each field is off-resonant
for optical excitation of level $b$, but the difference of the field
frequencies is close to that of the Raman transition. The control and input
fields have a relative propagation vector $\mathbf{K}$ that leads to a Doppler
shift $\mathbf{K\cdot v}$ associated with the two-photon Raman transition. As
a consequence the bandwidth that can be absorbed in this Raman process is on
the order of $1/Ku$, where $u$ is a characteristic atomic speed. Of critical
importance is that all frequency components of the signal field are depleted
in an identical fashion (no pulse distortion) if the bandwidth is less than
the inhomogeneous width. The medium is optically dense so the signal pulse is
totally attenuated. In other words, as a result of stimulated Raman
scattering, the signal pulse energy is totally transferred to the control
field. Each atom has a negligibly small population in state $\left\vert
{c}\right\rangle $ and the entire population stored in level $c$ is assumed to
be small as well, assuming the signal pulse is weak. The control field is
turned off following the depletion of the signal pulse. In contrast to the MK
protocol in which an optical coherence is created in the first stage, the
signal field is transferred directly to a Raman coherence in our protocol that
is immune to spontaneous emission decay.

\subsubsection{Stage 2}

The Raman coherence dephases following excitation as a result of the
inhomogeneous broadening. As in a photon echo experiment, this dephasing can
be reversed by the application of a second pair of pulses. However, if we were
to use a two-pulse echo process, the Raman coherence excited by the signal
would be affected by velocity changing collisions over the \textit{entire}
storage time between the two pulses. Instead, if we use a three-pulse echo
configuration, the effect of collisions during most of the storage time can be
suppressed. The second pulse pair, in effect, freezes the Doppler phase
created by the first pulse pair. As in the MK protocol, this phase reversal
must be carried out with a Raman pulse that leaves the population in level $a$
unchanged. This is a crucial condition to ensure uniform illumination by the
control fields; if any control field is resonant with a transition originating
in level $a$, the field will be strongly absorbed in the optically dense
medium and unsuitable for this protocol \cite{pop}. All these requirements are
satisfied if one applies a Raman $\pi$ pulse between levels $c$ and $d$,
following the excitation of the $a-c$ Raman coherence. The relative
$\mathbf{K}$ vector of the two fields is the same as that in stage 1, so the
net effect of the $\pi$ pulse is to convert the $a-c$ to $a-d$ coherence,
while freezing the Doppler phase evolution of this Raman coherence, just as in
a stimulated photon echo. The net result is that the original pulse
information is now stored in a Raman coherence that is, for the most part,
"protected" from the effects of velocity-changing collisions.

\subsubsection{Stage 3}

To prepare the system for the retrieval stage, a $\pi$ pulse is sent into the
medium at some later time to restore the $a-c$ coherence. This Raman pulse has
its relative $\mathbf{K}$ vector reversed and prepares the atoms with a
spatial Raman coherence that allows for retrieval of the signal in stage 4.

\subsubsection{Stage 4}

A control pulse is sent into the sample that is identical to the initial
control pulse, but with its propagation vector \textit{reversed}. The first
three stages produce a Raman coherence that allows a field to build up in a
direction opposite to that of the input signal pulse. In other words, the
depletion of the signal field is reversed and the original signal pulse is
restored as it exits the sample propagating in a direction opposite to that of
the original signal pulse. In principle, the pulse can be restored with close
to 100\% fidelity. In the context of creating a functional quantum memory
device, splitting the phase reversal in two steps (stages 2 and 3) reduces the
waiting time between the read-out pulse and the emission of the restored
signal, making the retrieved data available more rapidly after the read-out decision.

We now describe in detail how each of these steps can be achieved.

\section{Storage step}

Storage corresponds to the mapping of the optically carried information into
atomic Raman coherence, accompanied by attenuation of the input signal field.

\subsection{Buildup of a Raman atomic superposition state}

The input signal is depleted by stimulated Raman scattering in a three-level
$\Lambda$-system. The weak pulse to be stored, combined with the control
field, resonantly excites the Raman transition $\left\vert {a}\right\rangle
-\left\vert {c}\right\rangle $. Both fields are tuned off resonance from the
optical transition to upper level $\left\vert {b}\right\rangle $. Storage is
performed into the superposition of the ground substates $\left\vert
{a}\right\rangle $ and $\left\vert {c}\right\rangle $. Since the atoms are
prepared initially in state $\left\vert {a}\right\rangle $, the medium is
transparent to the control field that uniformly illuminates all the active
atoms. Assuming that the control field is constant during the signal pulse, we
write the electric field of the control field as the plane wave
\begin{equation}
E_{2}(\mathbf{r},t)=\mathcal{A}_{2}e^{i\mathbf{k}_{2}.\mathbf{r}-i\omega_{2}%
t}+c.c \label{cont}%
\end{equation}
where the envelope $\mathcal{A}_{2}$ is a time- and space-independent
parameter. The control field wave vector and frequency are denoted
$\mathbf{k}_{2}$ and $\omega_{2}$, respectively.

The signal pulse Rayleigh range is assumed to be much larger than the storage
material length $L$ to insure that its diameter does not vary significantly as
it propagates in the medium. The electric field of the signal field can be
expressed as
\begin{equation}
E_{1}(\mathbf{r},t)=\mathcal{A}_{1}(\mathbf{r},t)e^{i\mathbf{k}_{1}%
.\mathbf{r}-i\omega_{1}t}+c.c., \label{sigf}%
\end{equation}
where $\mathcal{A}_{1}(\mathbf{r},t)$ is the envelope, $\mathbf{k}_{1}$ the
propagation vector, and $\omega_{1}$ the carrier frequency of this field. The
spatial dependence of $\mathcal{A}_{1}(\mathbf{r},t)$ reflects the radial
distribution of the field and its attenuation along direction $\mathbf{k}_{1}%
$. When $L/c$ is not much smaller than the pulse duration, retardation also
contributes to the field envelope spatial dependence. The Rabi frequencies
associated with the signal and control fields are denoted by
\begin{subequations}
\label{rabi}%
\begin{align}
\Omega_{1}(\mathbf{r},t)  &  =-\mu_{ba}\mathcal{A}_{1}(\mathbf{r}%
,t)/\hbar;\label{rabia}\\
\Omega_{2}  &  =-\mu_{bc}\mathcal{A}_{2}/\hbar, \label{rabib}%
\end{align}
where $\mu_{ba}$ and $\mu_{bc}$ are optical dipole moment matrix elements. It
is assumed that $k_{1}\approx k_{2}\equiv k$.

In perturbation theory with the population of level $a$ set equal to unity,
the coupled equations for the optical and Raman coherences are:
\end{subequations}
\begin{subequations}
\label{eqs}%
\begin{align}
\dot{\tilde{\rho}}_{ab;m}  &  =(i\Delta_{1}-\gamma_{ab})\tilde{\rho}%
_{ab;m}+i\Omega_{1}^{\ast}\left[  \mathbf{r}_{m}(t),t\right]  e^{-i\mathbf{k}%
_{1}.\mathbf{r}_{m}(t)}+i\tilde{\rho}_{ac;m}\Omega_{2}^{\ast}e^{-i\mathbf{k}%
_{2}.\mathbf{r}_{m}(t)}\\
\dot{\tilde{\rho}}_{ac;m}  &  =[i(\Delta_{1}-\Delta_{2})-\gamma_{ac}%
]\tilde{\rho}_{ac;m}+i\tilde{\rho}_{ab;m}\Omega_{2}e^{i\mathbf{k}%
_{2}.\mathbf{r}_{m}(t)} \label{Raman_coherence_0}%
\end{align}
where $\Delta_{1}$ and $\Delta_{2}$ are the atom-field detunings for each
optical transition, $\gamma_{ab}$ is the decay rate for the $a-b$ coherence,
$\gamma_{ac}$ is the decay rate for the $a-c$ coherence, $\tilde{\rho}%
_{ab;m}=\rho_{ab;m}e^{-i\omega_{1}t}$ and $\tilde{\rho}_{ac;m}=\rho
_{ac;m}e^{-i(\omega_{1}-\omega_{2})t}$. These equations give the time
evolution of the density matrix elements for atom $m$, located at
$\mathbf{r}_{m}(t)$ at time $t$. The manner in which the spatial phases of the
field are imprinted on the atoms is readily apparent in Eqs. (\ref{eqs}).
Under the assumption that $\Omega_{1},\Omega_{1}^{-1}d\Omega_{1}%
/dt,\gamma_{ab},ku<<\Delta_{1}$, where $u$ is the most probable atomic speed,
the optical coherence adiabatically follows the field variations and can be
written as:
\end{subequations}
\begin{equation}
\tilde{\rho}_{ab;m}=-\Omega_{1}^{\ast}\left[  \mathbf{r}_{m}(t),t\right]
e^{-i\mathbf{k}_{1}.\mathbf{r}_{m}(t)}/\Delta_{1}-\tilde{\rho}_{ac;m}%
\Omega_{2}^{\ast}e^{-i\mathbf{k}_{2}.\mathbf{r}_{m}(t)}/\Delta_{1}.
\label{optical_coherence}%
\end{equation}
Substituting this expression into Eq. \ref{Raman_coherence_0}, one obtains
\begin{equation}
\dot{\tilde{\rho}}_{ac;m}=\left[  i\left(  \Delta_{1}-\Delta_{2}%
-\frac{\left\vert \Omega_{2}\right\vert ^{2}}{\Delta_{1}}\right)  -\gamma
_{ac}\right]  \tilde{\rho}_{ac;m}-i\frac{\Omega_{1}^{\ast}\left[
\mathbf{r}_{m}(t),t\right]  \Omega_{2}e^{i\mathbf{K}.\mathbf{r}_{m}(t)}%
}{\Delta_{1}}%
\end{equation}
where
\[
\mathbf{K}=\mathbf{k}_{2}-\mathbf{k}_{1}.
\]
The detuning $\Delta_{1}-\Delta_{2}$ can be adjusted to cancel the light shift
$\left\vert \Omega_{2}\right\vert ^{2}/\Delta_{1}$. Finally, the Raman
coherence can be expressed as:
\begin{equation}
\tilde{\rho}_{ac;m}(t)=-i\frac{\Omega_{2}}{\Delta_{1}}\int_{-\infty}%
^{t}dt^{\prime}\Omega_{1}^{\ast}\left[  \mathbf{r}_{m}(t^{\prime}),t^{\prime
}\right]  e^{i\mathbf{K}.\mathbf{r}_{m}(t^{\prime})}%
\end{equation}
where it has been assumed that any decay of $\tilde{\rho}_{ac;m}$ during the
signal pulse can be neglected. The density matrix element can be expressed in
terms of the atom position at time $t$. Indeed, if collisions do not change
the atomic velocity $\mathbf{v}_{m}$ during the signal pulse, the position at
$t^{\prime}$ can be expressed as $\mathbf{r}_{m}(t^{\prime})=\mathbf{r}%
_{m}(t)-\mathbf{v}_{m}(t-t^{\prime})$, so that:
\begin{equation}
\tilde{\rho}_{ac;m}(t)=-i\frac{\Omega_{2}}{\Delta_{1}}e^{i\mathbf{K}%
.\mathbf{r}_{m}(t)}\int_{-\infty}^{t}dt^{\prime}\Omega_{1}^{\ast}\left[
\mathbf{r}_{m}(t)-\mathbf{v}_{m}(t-t^{\prime}),t^{\prime}\right]
e^{-i\mathbf{K}.\mathbf{v}_{m}(t-t^{\prime})} \label{Raman_coherence}%
\end{equation}

\subsection{Stimulated Raman scattering of the input signal}

It is assumed that, in the absence of the control field, the scattering of the
probe field is negligible owing to the large detuning $\Delta_{1}$. As a
consequence, the contribution to the polarization resulting from the first
term in Eq. (\ref{optical_coherence}) can be neglected. On the other hand, the
control field intensity is sufficiently large to allow for significant
stimulated Raman scattering, resulting in a loss of signal field intensity as
the signal field propagates in the medium. The Raman contribution to
$\tilde{\rho}_{ab;m}$ is given by the second term on the right hand side of
Eq. (\ref{optical_coherence}). To determine the modification of the signal
field, we need to calculate the polarization associated with the $\tilde{\rho
}_{ab;m}$ coherence. We write the macroscopic polarization as
\[
P(\mathbf{r},t)=P_{+}(\mathbf{r},t)e^{i\mathbf{k}_{1}.\mathbf{r-}i\omega_{1}%
t}+P_{+}^{\ast}(\mathbf{r},t)e^{-\left(  i\mathbf{k}_{1}.\mathbf{r-}%
i\omega_{1}t\right)  }%
\]

In going over to a macroscopic polarization, we assume that, at any position
$\mathbf{r}$ in the medium, one can define a slice of thickness $l<<2\pi/k$ in
the $\mathbf{k}_{1}$ direction, containing many atoms. The macroscopic
polarization at $(\mathbf{r},t)$ is obtained by combining the contributions
from all the atoms within the slice. Those atoms satisfy the condition
$\left\vert \mathbf{\hat{k}}_{1}.\mathbf{s}_{m}(t)\right\vert \leq l/2$, where
$\mathbf{\hat{k}}_{1}$ is a unit vector in the $\mathbf{k}_{1}$ direction and
$\mathbf{s}_{m}(t)=\mathbf{r}-\mathbf{r}_{m}(t)$. Therefore the positive
frequency component $P_{+}(\mathbf{r},t)$ is given by
\begin{equation}
P_{+}(\mathbf{r},t)=\frac{\mu_{ab}}{\delta V}e^{-i\mathbf{k}_{1}.\mathbf{r}%
}\displaystyle\sum_{\substack{m\\\left\vert \mathbf{\hat{k}}_{1}%
.\mathbf{s}_{m}(t)\right\vert \leq l/2}}\tilde{\rho}_{ba;m}\left(
\mathbf{r},t\right)  , \label{pol}%
\end{equation}
where $\delta V$ represents the slice volume. Combining Eqs.
(\ref{optical_coherence}), (\ref{Raman_coherence}), (\ref{pol}), and
(\ref{rabia}), we find
\begin{equation}
P_{+}(\mathbf{r},t)=i\frac{\left\vert \mu_{ab}\right\vert ^{2}\left\vert
\Omega_{2}\right\vert ^{2}}{\delta V\hbar\Delta_{1}^{2}}\displaystyle\sum
_{\substack{m\\\left\vert \mathbf{\hat{k}}_{1}.\mathbf{s}_{m}(t)\right\vert
\leq l/2}}\int_{-\infty}^{t}dt^{\prime}\mathcal{A}_{1}\left[  \mathbf{r}%
-\mathbf{v}_{m}(t-t^{\prime}),t^{\prime}\right]  e^{i\mathbf{K}.\mathbf{v}%
_{m}(t-t^{\prime})}. \label{pp}%
\end{equation}
As was noted above, the contribution to $\tilde{\rho}_{ba;m}\left(
\mathbf{r},t\right)  $ from the first term in Eq. (\ref{optical_coherence})
has been neglected since it adiabatically follows the field and vanishes for
times greater than the pulse duration.

The atoms are uniformly distributed in space, with density $N$, and their
normalized velocity distribution is represented by $W(\mathbf{v})$. Replacing
the discrete sum by an integral, according to
\[
\frac{1}{\delta V}\displaystyle\sum_{\substack{m\\\left\vert \mathbf{\hat{k}%
}_{1}.\mathbf{s}_{m}(t)\right\vert \leq l/2}}\rightarrow N\int d^{3}%
vW(\mathbf{v}),
\]
enables us to transform Eq. (\ref{pp}) into
\begin{equation}
P_{+}(\mathbf{r},t)=i\frac{\left\vert \mu_{ab}\right\vert ^{2}\left\vert
\Omega_{2}\right\vert ^{2}}{\hbar\Delta_{1}^{2}}N\int d^{3}vW(\mathbf{v}%
)\int_{-\infty}^{t}dt^{\prime}\mathcal{A}_{1}\left[  \mathbf{r}-\mathbf{v}%
(t-t^{\prime}),t^{\prime}\right]  e^{i\mathbf{K}.\mathbf{v}(t-t^{\prime})}.
\end{equation}
Provided the input signal spectral width $\delta_{s}$ is smaller than $Ku$,
$\mathcal{A}_{1}\left[  \mathbf{r}-\mathbf{v}(t-t^{\prime}),t^{\prime}\right]
$ can be taken out of the integral over $t^{\prime}$ and evaluated at
$t^{\prime}=t$. In this limit the polarization $P_{+}(\mathbf{r},t)$ reduces
to:
\begin{subequations}
\label{initial_polarization}%
\begin{align}
P_{+}(\mathbf{r},t)  &  =i\frac{\left\vert \mu_{ab}\right\vert ^{2}\left\vert
\Omega_{2}\right\vert ^{2}}{\hbar\Delta_{1}^{2}}\mathcal{A}_{1}(\mathbf{r}%
,t)N\int dvW(v)\int_{0}^{\infty}d\tau e^{iKv\tau}\\
&  =\frac{i}{k}\frac{N\pi\left\vert \mu_{ab}\right\vert ^{2}W(0)}{\hbar}%
\frac{k\left\vert \Omega_{2}\right\vert ^{2}}{K\Delta_{1}^{2}}\mathcal{A}%
_{1}(\mathbf{r},t)
\end{align}
where $W(v)$ represents the one-dimensional velocity distribution.

This is the key result of this section. Owing to the large inhomogeneous
width, the polarization is proportional to the field amplitude and depends
locally on this amplitude. In other words, the polarization does not depend on
the value of the field amplitude at earlier times as it would in the case of
homogeneous broadening. The electric susceptibility $\chi_{R}$ is defined by
$P_{+}(\mathbf{r},t)=\epsilon_{0}\chi_{R}\mathcal{A}_{1}(\mathbf{r},t)$ with
the intensity absorption coefficient $\alpha_{R}$ given by $\alpha
_{R}=k\mathrm{Im}(\chi_{R})$, which leads to
\end{subequations}
\begin{equation}
\alpha_{R}=\frac{k\left\vert \Omega_{2}\right\vert ^{2}}{K\Delta_{1}^{2}%
}\alpha_{0} \label{Raman_absorption}%
\end{equation}
where $\alpha_{0}$ represents the linear absorption coefficient on the
inhomogeneously broadened transition $a-b$. Provided $\delta_{s}<Ku$, the
signal propagates without distortion, which implies that all the signal
spectral components are uniformly attenuated and stored in the atomic
ensemble. The spectral components are mapped into Raman coherence in atoms
whose velocities span an interval of order $\delta_{s}/K$ along the direction
$\mathbf{K=k}_{2}-\mathbf{k}_{1}$. The parameter $K$ should be adjusted in
such a way that $\delta_{s}/Ku$ is larger than unity to provide a sufficiently
large bandwidth to store the signal pulse, but not too large, since the signal
depletion and Raman storage varies inversely with $K$.

\section{Freezing the Doppler phase}

The input signal illuminates the storage medium during a time interval
centered at $t_{1}$. The control field is turned off following the signal
pulse. At some later time, the Raman coherence, given by Eq.
(\ref{Raman_coherence}), can be expressed as:
\begin{equation}
\tilde{\rho}_{ac;m}(t)=-i\mathrm{e}^{i\mathbf{K}.\mathbf{r}_{m}(t)-\left(
i\mathbf{K}.\mathbf{v}_{m}+\gamma_{ac}\right) (t-t_{1})}\mathcal{R}%
_{m}\label{Raman_coherence_2}%
\end{equation}
where
\begin{equation}
\mathcal{R}_{m}=\frac{\Omega_{2}}{\Delta_{1}}\int_{-\infty}^{\infty}%
d\tau\Omega_{1}^{\ast}\left[  \mathbf{r}_{m}(t_{1}),t_{1}+\tau\right]
\mathrm{e}^{i\mathbf{K}.\mathbf{v}_{m}\tau} \label{rm}%
\end{equation}
and it was assumed that $\Omega_{1}\left[  \mathbf{r}-\mathbf{v}\tau
_{p},t\right]  \approx\Omega_{1}\left[  \mathbf{r},t\right]  $, where
$\tau_{p}$ is the signal pulse duration. There is a build-up of Doppler phase
associated with the Raman coherence that grows linearly as a function of time
following the interaction with the signal pulse. Although this phase could be
reversed at a later time, it is best to nip it in the bud to prevent any
deterioration from velocity-changing collisions. To accomplish this task one
can send in a Raman $\pi$ pulse having the same $\mathbf{K}$ vector that
transfers the amplitude from state $\left\vert {c}\right\rangle $ to an
auxiliary level $\left\vert {d}\right\rangle $, or, equivalently, converts
coherence $\tilde{\rho}_{ac}$ into $\tilde{\rho}_{ad}$.%
\begin{figure}
[ptb]
\begin{center}
\includegraphics[height=4.0568in,width=4.5307in]
{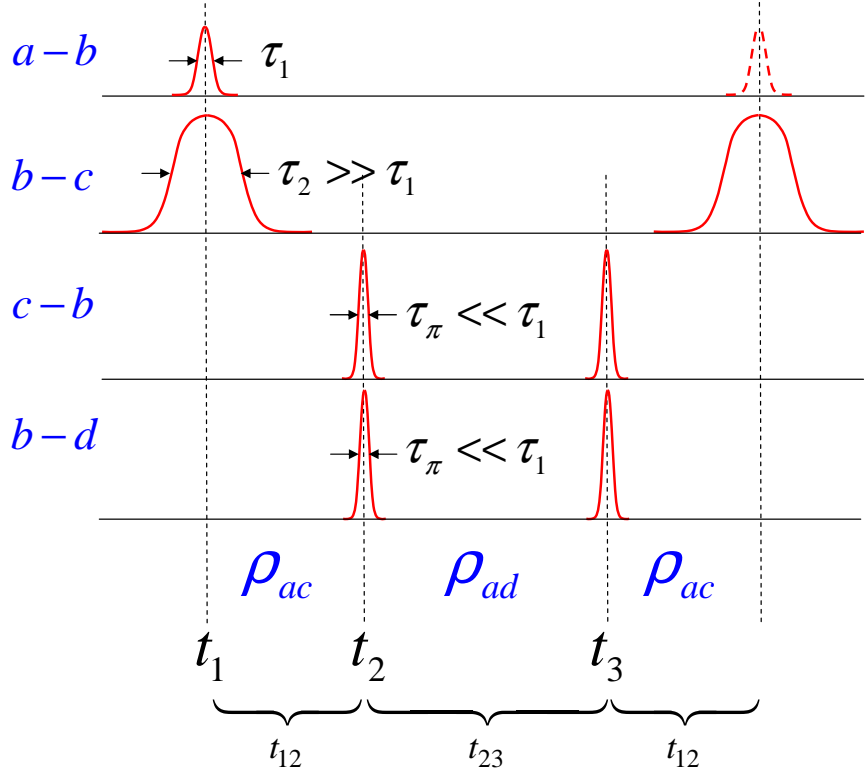}
\caption{(Color online) Schematic representation of the timing sequence for
the entire protocol, showing the relative duration of the different pulses,
and the various atomic quantities involved at the different stages.}%
\label{fig2a}%
\end{center}
\end{figure}

Let $\Omega_{3}(t)$ and $\Omega_{4}(t)$ denote the Rabi frequencies on
transitions $\left\vert {c}\right\rangle -\left\vert {b}\right\rangle $ and
$\left\vert {b}\right\rangle -\left\vert {d}\right\rangle $, respectively. The
radial extension of these fields is assumed to be larger than that of the
signal pulse. Moreover, these fields are not attenuated as they propagate
through the storage material since interact with quasi-empty levels. Hence
their Rabi frequency space dependence can be omitted. Both fields are detuned
by the same amount $\Delta$ from resonance with their respective assigned
transition. Therefore the $\left\vert {c}\right\rangle -\left\vert
{d}\right\rangle $ two-photon transition is resonantly excited. The Rabi
frequency of the equivalent two-level system is given by $\Omega_{3}%
(t)\Omega_{4}^{\ast}(t)/\Delta$. The two-level approximation holds provided
$\left\vert \tilde{\Omega}_{3,4}(\Delta\pm\delta_{s})\right\vert ^{2}<<1$,
where $\tilde{\Omega(\Delta)}$ represents the time-to-frequency Fourier
transform of $\Omega(t)$. Of course one must take care that none of those
fields can excite state $\left\vert {a}\right\rangle ,$ where all the atomic
population is concentrated. With these assumptions, one finds that $\rho_{ac}$
and $\rho_{ad}$ evolve according to
\begin{subequations}
\label{pi_pulse}%
\begin{align}
\dot{\tilde{\rho}}_{ad;m} &  =i\frac{\left\vert \Omega_{\pi}\right\vert ^{2}%
}{\Delta}\mathrm{e}^{-i\phi_{m}(t)}\tilde{\rho}_{ac;m}-\gamma_{ad}\tilde{\rho
}_{ad;m}\\
\dot{\tilde{\rho}}_{ac;m} &  =i\frac{\left\vert \Omega_{\pi}\right\vert ^{2}%
}{\Delta}\mathrm{e}^{i\phi_{m}(t)}\tilde{\rho}_{ad;m}-\gamma_{ac}\tilde{\rho
}_{ac;m},
\end{align}
where $\tilde{\rho}_{ad;m}=\rho_{ad;m}\mathrm{e}^{-i(\omega_{3}-\omega_{4})t}%
$, $\phi_{m}(t)=(\mathbf{k}_{3}-\mathbf{k}_{4}).\mathbf{r}_{m}(t)$ and, for
simplicity, we have set $\Omega_{3}(t)=\Omega_{4}(t)=\Omega_{\pi}(t)$ and
neglected a light shift that is the same for levels $c$ and $d$. The pulses,
having duration $\tau_{\pi}$, are applied at a time centered about $t=$
$t_{2}$ and their duration is assumed to be sufficiently short, $\delta
_{s}\tau_{\pi}<<1$, to resonantly excite all the atoms that were excited by
the signal pulse (recall that atoms in the velocity range $\left\vert
\mathbf{K\cdot v}\right\vert \leq\left\vert \delta_{s}\right\vert $ are
excited in stage 1). The time-diagram of the whole protocol is displayed in
Fig. \ref{fig2a}, showing the relative duration of the different pulses.

In terms of the coherence components at time $t_{2}^{-}$ just before the Raman
$\pi$ pulse is applied, Eq. (\ref{pi_pulse}) can be solved, for $t>t_{2}$,
as:
\end{subequations}
\begin{subequations}
\label{Doppler_freeze}%
\begin{align}
\tilde{\rho}_{ac;m}(t)  &  =\mathrm{e}^{-\gamma_{ac}(t-t_{2})}\cos\left(
\frac{\Theta}{2} \right) \tilde{\rho}_{ac;m}(t_{2}^{-})+i\mathrm{e}^{i\phi
_{m}(t_{2})-\gamma_{ad}(t-t_{2})}\sin\left(  \frac{\Theta}{2} \right)
\tilde{\rho}_{ad;m}(t_{2}^{-})\\
\tilde{\rho}_{ad;m}(t)  &  =i\mathrm{e}^{-i\phi_{m}(t_{2})-\gamma_{ad}%
(t-t_{2})}\sin\left( \frac{\Theta}{2} \right)  \tilde{\rho}_{ac;m}(t_{2}%
^{-})+\mathrm{e}^{-\gamma_{ad}(t-t_{2})}\cos\left(  \frac{\Theta}{2} \right)
\tilde{\rho}_{ad;m}(t_{2}^{-}),
\end{align}
where $\Theta=2\int dt\left\vert \Omega_{\pi}(t)\right\vert ^{2}/\Delta$,
assuming that $\gamma_{ac}\tau_{\pi},\gamma_{ad}\tau_{\pi}<<1$. Total
conversion from $\rho_{ac}$ to $\rho_{ad}$ requires that $\Theta=\pi$. With
initial conditions given by Eq. (\ref{Raman_coherence_2}) and $\tilde{\rho
}_{ad;m}(t_{2}^{-})=0$ one finds that
\end{subequations}
\begin{equation}
\tilde{\rho}_{ad;m}(t)=\mathrm{e}^{i(\mathbf{k}_{4}-\mathbf{k}_{3}
+\mathbf{K}).\mathbf{r}_{m}(t_{2})-\gamma_{ad}(t-t_{2})}\mathrm{e}^{-\left(
i\mathbf{K}.\mathbf{v}_{m}+\gamma_{ac}\right)  t_{12}}\mathcal{R}%
_{m},\label{stopped_dephasing}%
\end{equation}
where $t_{ij}$ is the time interval between the $j$th and $i$th pairs of
pulses. With $\mathbf{k}_{3}-\mathbf{k}_{4}=\mathbf{K}$, there is no build-up
of Doppler phase following this second Raman pulse \cite{doppler} - the
Doppler phase has been frozen.

\section{Signal recovery}

To prepare for signal retrieval at some later time, one sends in another Raman
$\pi$ pulse, centered at time $t=t_{3}$, to restore the $\tilde{\rho}_{ac;m}$
coherence. This Raman pulse consists of two fields having propagation vectors
$\mathbf{k}_{4}^{\prime}$ and $\mathbf{k}_{3}^{\prime}$ that drive the
$\left\vert {d}\right\rangle -\left\vert {b}\right\rangle $ and $\left\vert
{b}\right\rangle -\left\vert {c}\right\rangle $ transitions, respectively. The
atom-field dynamics is again described by Eqs.(\ref{pi_pulse}), with $\phi
_{m}(t)$ replaced by $\phi_{m}^{\prime}(t)=(\mathbf{k}_{3}^{\prime}%
-\mathbf{k}_{4}^{\prime}).\mathbf{r}_{m}(t).$ To reverse the Doppler phase
acquired in the time interval $t_{12}$, we choose the propagation vectors such
that $\mathbf{k}_{3}^{\prime}-\mathbf{k}_{4}^{\prime}=-\mathbf{K}$\textbf{,
}which leads to a Raman coherence for $t>t_{3}$ given by
\begin{align}
\tilde{\rho}_{ac;m}(t) &  =i\mathrm{e}^{i\phi_{m}^{\prime}(t_{3}%
)-i\mathbf{K}.\mathbf{v}_{m}t_{12}-\gamma_{ac}(t-t_{3}+t_{12})-\gamma
_{ad}t_{23}}\mathcal{R}_{m}\nonumber\\
&  =i\mathrm{e}^{-i\mathbf{K}.\mathbf{r}_{m}(t)+i\mathbf{K}.\mathbf{v}%
_{m}(t-t_{3})-i\mathbf{K}.\mathbf{v}_{m}t_{12}-\gamma_{ac}(t-t_{3}%
+t_{12})-\gamma_{ad}t_{23}}\mathcal{R}_{m}.\label{rac}%
\end{align}

To restore the signal one directs a quasi-monochromatic control field having
frequency $\omega_{2}$ into the medium in the backward direction, with
$\mathbf{k}_{2}^{\prime\prime}=-\mathbf{k}_{2}$. This field should be switched
on somewhat before time $t=t_{3}+t_{12}$, with the same Rabi frequency
$\Omega_{2}$ as the initial coupling beam. The restored field can be written
as
\[
E(\mathbf{r},t)=\mathcal{A}(\mathbf{r},t)\mathrm{e}^{i\mathbf{k}_{1}%
^{\prime\prime}.\mathbf{r-}i\omega_{1}t}+\mathrm{c}.\mathrm{c}.
\]
We now argue that, only if $\mathbf{k}_{1}^{\prime\prime}=-\mathbf{k}_{1}$,
can a phase matched signal be generated. To see this, we return to Eq.
(\ref{rac}) and analyze the phase factor in this equation. Although
$\mathcal{R}_{m}$ contains a Doppler phase factor [see Eq. (\ref{rm})], it is
of order $Ku\tau$, where $\tau$ is the signal pulse duration, and is small
compared to the other Doppler phases appearing in Eq. (\ref{rac}), since we
assume that $t_{12}\gg\tau$. When the control field is sent into the medium,
the Raman coherence (\ref{rac}) gives rise to an optical coherence on the
$a-b$ transition that is responsible for the generation of the restored field.
>From Eq. (\ref{optical_coherence}), one can deduce that the optical coherence
$\tilde{\rho}_{ab;m}$ varies as
\begin{align*}
&  \mathrm{e}^{i\mathbf{k}_{2}.\mathbf{r}_{m}(t)-i\mathbf{K}.\mathbf{r}%
_{m}(t)+i\mathbf{K}.\mathbf{v}_{m}(t-t_{3})-i\mathbf{K}.\mathbf{v}_{m}t_{12}%
}\\
&  =\mathrm{e}^{i\mathbf{k}_{1}.\mathbf{r}_{m}(t)+i\mathbf{K}.\mathbf{v}%
_{m}(t-t_{3}-t_{12})}%
\end{align*}
This expression implies that a phase matched signal can propagate only in the
$\mathbf{k}_{1}^{\prime\prime}=-\mathbf{k}_{1}$ direction and that this signal
can be nonvanishing (owing to the average over atomic velocities) only for
times $t\approx t_{3}+t_{12}$. Thus, in what follows we assume that the
restored signal field has propagation vector $\mathbf{k}_{1}^{\prime\prime
}=-\mathbf{k}_{1}$.

The existence of the restored field must be taken into account in the
expression for the Raman coherence. Indeed, this field also combines with the
control field to drive Raman transitions between levels $a$ and $c$. Adding
this contribution to Eq. (\ref{rac}), one obtains the Raman coherence
\begin{equation}%
\begin{split}
\tilde{\rho}_{ac;m}(t)  &  =i\mathrm{e}^{-i\mathbf{K}.\mathbf{r}%
_{m}(t)+i\mathbf{K}.\mathbf{v}_{m}(t-t_{3}-t_{12})-2\gamma_{ac}t_{12}%
-\gamma_{ad}t_{23}}\mathcal{R}_{m}\\
&  +i\frac{\mu_{ab}\Omega_{2}}{\hbar\Delta_{1}}\int_{-\infty}^{t}dt^{\prime
}\mathcal{A}\left[  \mathbf{r}_{m}(t^{\prime}),t^{\prime}\right]
\mathrm{e}^{i\mathbf{K^{\prime\prime}}.\mathbf{r}_{m}(t^{\prime})}.
\end{split}
\label{rephased_coherence}%
\end{equation}

Substituting this equation into Eq. (\ref{optical_coherence}) one arrives at
an expression for the optical coherence,
\begin{equation}%
\begin{split}
\tilde{\rho}_{ab;m}(t) &  =i\frac{\mu_{ab}\left\vert \Omega_{2}\right\vert
^{2}}{\hbar\Delta_{1}^{2}}\mathrm{e}^{i\mathbf{k}_{1}.\mathbf{r}_{m}(t)}\\
&  \times\left(  \mathrm{e}^{-\gamma_{ac}\left(  t-t_{3}-t_{12}\right)
-\gamma_{ad}t_{23}}\int_{-\infty}^{\infty}d\tau\mathcal{A}_{1}^{\ast}\left[
\mathbf{r}_{m}(t_{1}),t_{1}+\tau\right]  \mathrm{e}^{i\mathbf{K}%
.\mathbf{v}_{m}(t-t_{3}-t_{12}+\tau)}\right.  \\
&  \left.  -\int_{-\infty}^{t}dt^{\prime}\mathcal{A}^{\ast}\left[
\mathbf{r}_{m}(t^{\prime}),t^{\prime}\right]  \mathrm{e}^{i\mathbf{K}%
.\mathbf{v}_{m}(t-t^{\prime})}\right)  ,
\end{split}
\label{final_optical_coherence}%
\end{equation}
that can be used to calculate the macroscopic polarization density, which is
written as
\[
P_{s}(\mathbf{r},t)=P_{+s}(\mathbf{r},t)\mathrm{e}^{-i\mathbf{k}%
_{1}.\mathbf{r-}i\omega_{1}t}+P_{+s}^{\ast}(\mathbf{r},t)\mathrm{e}^{-\left(
-i\mathbf{k}_{1}.\mathbf{r-}i\omega_{1}t\right)  },
\]
incorporating the fact that the signal is phase-matched in the $-\mathbf{k}%
_{1}$ direction only. The polarization density is comprised of two components.
One of them represents the source term that gives rise to the restored signal.
This contribution originates from the first term on the right hand side of Eq.
(\ref{final_optical_coherence}) and is given by
\begin{equation}%
\begin{split}
P_{+s}^{(1)}(\mathbf{r},t) &  =-i\frac{\left\vert \mu_{ab}\right\vert
^{2}\left\vert \Omega_{2}\right\vert ^{2}N}{\hbar\Delta_{1}^{2}}%
\mathrm{e}^{-\gamma_{ac}\left(  t-t_{3}+t_{12}\right)  -\gamma_{ad}t_{23}}\\
&  \times\int d\mathbf{v}W(\mathbf{v})\int_{-\infty}^{\infty}d\tau
\mathcal{A}_{1}\left[  \mathbf{r}(t_{1}),t_{1}+\tau\right]  \mathrm{e}%
^{-i\mathbf{K}.\mathbf{v}(t-t_{3}-t_{12}+\tau)},
\end{split}
\end{equation}
where $\mathcal{A}_{1}\left[  \mathbf{r}(t_{1}),t^{\prime}\right]
\approx\mathcal{A}_{1}\left[  \mathbf{r}-\mathbf{v}t_{13},t^{\prime}\right]
$. The polarization density depends on the field experienced by the
participating atoms at their positions when they interacted with the signal field.

For the field to be restored, the condition $\mathcal{A}_{1}\left[
\mathbf{r}_{m}(t_{1}),t^{\prime}\right]  \approx\mathcal{A}_{1}\left[
\mathbf{r},t^{\prime}\right]  $ must hold. This condition is valid provided
the distance travelled by the atoms in time $t_{13}$ is much smaller than the
beam diameter, the spatial width of the input signal envelope, and the
absorption length $\alpha_{R}^{-1}$. Since $\delta_{s}<<Ku$, one can take
$\mathcal{A}_{1}\left[  \mathbf{r},t^{\prime}\right]  $ out of the integral
over $t^{\prime}$ and evaluate it at time $t^{\prime}=t_{1}-(t-t_{3}-t_{12})$,
which leads to
\begin{equation}%
\begin{split}
P_{+s}^{(1)}(\mathbf{r},t) &  =-i\frac{\left\vert \mu_{ab}\right\vert
^{2}\left\vert \Omega_{2}\right\vert ^{2}N}{\hbar\Delta_{1}^{2}}%
\mathrm{e}^{-\gamma_{ac}\left(  t-t_{3}+t_{12}\right)  -\gamma_{ad}t_{23}}\\
&  \times\mathcal{A}_{1}\left[  \mathbf{r},t_{1}-(t-t_{3}-t_{12})\right]  \int
dvW(v)\int_{-\infty}^{\infty}d\tau\mathrm{e}^{-iKv\tau}%
\end{split}
\end{equation}
Using Eqs. (\ref{initial_polarization}) and (\ref{Raman_absorption}), one can
re-express this in terms of $\alpha_{R}$ as%
\begin{equation}
P_{+s}^{(1)}(\mathbf{r},t)=-2i\epsilon_{0}\frac{\alpha_{R}}{k}\mathcal{A}%
_{1}\left[  \mathbf{r},t_{1}-(t-t_{3}-t_{12})\right]  \mathrm{e}%
^{-2\gamma_{ac}t_{12}-\gamma_{ad}t_{23}}%
\end{equation}
The amplitude $\mathcal{A}_{1}\left[  \mathbf{r},t_{1}-(t-t_{3}-t_{12}%
)\right]  $ is nonvanishing only for $t\approx t_{3}+t_{12},$ reflecting the
fact that Doppler rephasing occurs only for such times. We have used this fact
to set $e^{-\gamma_{ac}(t-t_{3}-t_{12})}\approx e^{-2\gamma_{ac}t_{12}}.$

The input signal was attenuated as it traveled through the storage medium.
Taking into account the propagation and attenuation of the input field
$A_{1}\left[ \mathbf{r},t\right] $, we can write $P_{+}^{(1)}(r,t)$ in terms
of the input field at $z=0$ as
\begin{equation}
P_{+s}^{(1)}(z,t)=-2i\epsilon_{0}\frac{\alpha_{R}}{k}\mathcal{A}_{1}\left[
0,t_{1}-(t-t_{3}-t_{12})-z/c\right]  \mathrm{e}^{-\alpha_{R}z/2}%
\mathrm{e}^{-2\gamma_{ac}t_{12}-\gamma_{ad}t_{23}}, \label{P1}%
\end{equation}
the $z$ axis being directed along $\mathbf{k}_{1}$.

The contribution to the polarization density from the second term on the right
hand side of Eq. (\ref{final_optical_coherence}) corresponds simply to
depletion of the restored field as a result of Raman transitions from level
$a$ to $c$ . This component can be written as
\begin{equation}
P_{+s}^{(2)}(z,t)=i\epsilon_{0}\frac{\alpha_{R}}{k}\mathcal{A}(z,t).\label{P2}%
\end{equation}
Finally, the restored field is a solution of the linearized Maxwell equation:
\begin{equation}
-\partial_{z}\mathcal{A}(z,t)+\frac{1}{c}\partial_{t}\mathcal{A}%
(z,t)=i\frac{k}{2\epsilon_{0}}\left[  P_{+s}^{(1)}(z,t)+P_{+s}^{(2)}%
(z,t)\right]  \label{P3}%
\end{equation}
Making the change of variables $z^{\prime}=z$, $t^{\prime}=t+z/c$, and
combining Eqs. (\ref{P1}) and (\ref{P3}), one obtains
\begin{equation}
\partial_{z^{\prime}}\mathcal{A}^{\prime}(z^{\prime},t^{\prime})=-\alpha
_{R}\mathcal{A}_{1}\left[  0,t_{1}-(t^{\prime}-t_{3}-t_{12})\right]
\mathrm{e}^{-\alpha_{R}z^{\prime}/2}\mathrm{e}^{-2\gamma_{ac}t_{12}%
-\gamma_{ad}t_{23}}+\frac{1}{2}\alpha_{R}\mathcal{A}^{\prime}(z^{\prime
},t^{\prime}),
\end{equation}
where $\mathcal{A}^{\prime}(z^{\prime},t^{\prime})=\mathcal{A}(z^{\prime
},t^{\prime}-z^{\prime}/c)$. Solving this equation with the initial condition
$\mathcal{A}^{\prime}(L,t^{\prime})=0,$ we find
\begin{equation}
\mathcal{A}(z,t)=\mathcal{A}_{1}\left[  z,t_{1}-(t-t_{3}-t_{12})\right]
\mathrm{e}^{-2\gamma_{ac}t_{12}-\gamma_{ad}t_{23}}\left(  1-\mathrm{e}%
^{-\alpha_{R}(L-z)}\right)  .\label{restored_field}%
\end{equation}
For $\alpha_{R}L\gg1$, the pulse is totally restored at time $t=t_{3}+t_{12}$
, neglecting decay of the Raman coherences. The latter equation represents the
main result of the paper, showing the absence of distortion of the retrieved
signal and its variation as a function of the optical depth $\alpha_{R}L$.

One can also verify that the atoms return to their initial state as the signal
field is restored. Substituting Eq. (\ref{restored_field}) into Eq.
(\ref{rephased_coherence}) and using Eq. (\ref{rm}), one obtains
\begin{equation}%
\begin{split}
&  \tilde{\rho}_{ac;m}(t)=-i\frac{\mu_{ab}\Omega_{2}}{\hbar\Delta_{1}%
}\mathrm{e}^{-i\mathbf{K}.\mathbf{r}_{m}(t)+i\mathbf{K}.\mathbf{v}_{m}%
(t-t_{3}-t_{12})-\gamma_{ac}(2t_{23}+t_{12})}\\
&  \times\left(  \int_{-\infty}^{\infty}dt^{\prime}\mathcal{A}_{1}^{\ast
}(z,t^{\prime})\mathrm{e}^{i\mathbf{K}.\mathbf{v}_{m}(t^{\prime}-t_{1}%
)}-\left(  1-\mathrm{e}^{-\alpha_{R}(L-z)}\right)  \int_{t_{1}-(t-t_{3}%
-t_{12})}^{\infty}dt^{\prime}\mathcal{A}_{1}^{\ast}(z,t^{\prime}%
)\mathrm{e}^{i\mathbf{K}.\mathbf{v}_{m}(t^{\prime}-t_{1})}\right)  .
\end{split}
\end{equation}
For times $\left(  t-t_{3}-t_{12}\right)  >0$ the lower limit on the second
integral can be replaced by $-\infty$ and this term cancels the first
integral. In other words, the coherence vanishes once when the signal field is
restored, provided $\alpha_{R}L\gg1$ .

Following the emission of the restored signal pulse, there remains in the
medium a Raman coherence given by
\begin{equation}
\tilde{\rho}_{ac;m}(t)=-i\frac{\mu_{ab}\Omega_{2}}{\hbar\Delta_{1}}%
\mathrm{e}^{-i\mathbf{K}.\mathbf{r}_{m}(t)+i\mathbf{K}.\mathbf{v}_{m}%
(t-t_{3}-t_{12})-\gamma_{ac}(t_{23}+2t_{12})}\mathrm{e}^{-\alpha_{R}(L-z)}%
\int_{-\infty}^{\infty}dt^{\prime}\mathcal{A}_{1}^{\ast}(z,t^{\prime
})\mathrm{e}^{i\mathbf{K}.\mathbf{v}_{m}(t^{\prime}-t_{1})}.
\end{equation}
Comparing with expression with Eq. (\ref{Raman_coherence}), and noting that
the population in state $\left\vert {a}\right\rangle $ is approximately equal
to unity, one can conclude that the population left in state $\left\vert
{c}\right\rangle $ is
\begin{subequations}
\begin{align}
\rho_{cc;m}[(t_{3}+t_{12})^{+}]  &  =\left\vert \tilde{\rho}_{ac;m}(t_{1}%
^{+})\right\vert ^{2}\left[  1-\mathrm{e}^{-2\gamma_{ac}(t_{23}+2t_{12}%
)}\right]  +\left\vert \tilde{\rho}_{ac;m}[(t_{3}+t_{12})^{+}]\right\vert
^{2}\\
&  =\left\vert \tilde{\rho}_{ac;m}(t_{1}^{+})\right\vert ^{2}\left[
1-\mathrm{e}^{-2\gamma_{ac}(t_{23}+2t_{12})}\left(  1-\mathrm{e}^{-2\alpha
_{R}(L-z)}\right)  \right]  .
\end{align}
Finally, summing over $z$, one finds the fraction $\eta$ of the initially
excited atoms that is left in state $\left\vert {c}\right\rangle $ is given
by
\end{subequations}
\begin{equation}
\eta=1-\mathrm{e}^{-2\gamma_{ac}(t_{23}+2t_{12})}\left(  1-\mathrm{e}%
^{-\alpha_{R}L}\right)  .
\end{equation}

If decay of the Raman coherence decay can be neglected, $\eta=\mathrm{e}%
^{-\alpha_{R}L}$. During the storage process, a fraction $\eta_{in}%
=(1-\mathrm{e}^{-\alpha_{R}L})$ of the incoming signal radiation is used to
excite the atoms to level $c$ , the remaining signal passes through without
being scattered. From the part that is stored, a fraction is lost at
retrieval, even in the absence of relaxation. Indeed the restored field is
$(1-\mathrm{e}^{-\alpha_{o}L})$ times smaller than the incoming one, according
to Eq. (\ref{restored_field}). Therefore one recovers a fraction
$W_{out}/W_{in}=(1-\mathrm{e}^{-\alpha_{R}L})^{2}$ of the incoming energy. The
difference $\eta_{in}-W_{out}/W_{in}\cong\mathrm{e}^{-\alpha_{R}L}$
corresponds to the fraction $\eta$ of the excited atoms that is left in level
$c$. To summarize, with a finite length material, information is lost in equal
amounts at storage and retrieval, the storage loss associated with an
incomplete depletion of the input field and the retrieval loss associated with
excited state population $\rho_{cc}$ that is left in the medium.

\section{Discussion}

We have shown that nearly 100$\%$ recovery efficiency can be reached with the
proposed quantum storage Raman scheme. The memory bandwidth, determined by the
Raman transition Doppler broadening, can be tuned continuously from $0$ to
$2ku$. Since the atoms are not excited to the upper electronic level, the
control pulse duration is not limited by the upper level lifetime, but only by
the inverse signal bandwidth.

Our motivation in this work was to circumvent the time scale conditions
imposed by the short upper level lifetime, since it is not easy to produce
large area, large waist coherent pulses on the nanosecond time scale. The
pulses to be considered here, whether input signal or control pulses, will
most likely last for tens or hundreds of nanoseconds. This corresponds to a
spectral width much smaller than the Doppler width $ku$, typically of order
$10^{9}s^{-1}$ or higher. The available bandwidth of the Raman scheme is
determined by $Ku$, where $K/k$ can be expressed in terms of the angle
$\theta=(\mathbf{k}_{1},\mathbf{k}_{2})$ as $K/k=2\theta\sin(\theta/2)$.
According to Eq. (\ref{restored_field}), the Raman optical depth $\alpha_{R}L$
should be as large as possible. Since $\alpha_{R}L$ is inversely proportional
to $K/k$, the value of $K/k$ should be matched to the bandwidth of the input
pulse. For an input field spectral width of order $10$ MHz, we are led to a
nearly co-propagating configuration with $\theta$ of order 10 mrad.

The memory lifetime is limited by the atomic motion. As noticed in Section IV,
the distance travelled by the atoms during the entire process should be much
smaller than the beam diameter and the absorption length $\alpha_{R}^{-1}$.
With a typical average speed of a few 100m/s, a memory lifetime of a few tens
of microseconds limits the sample size to a few centimeters.

\begin{figure}
[ptb]
\begin{center}
\includegraphics[height=4.0577in,width=2.9758in]
{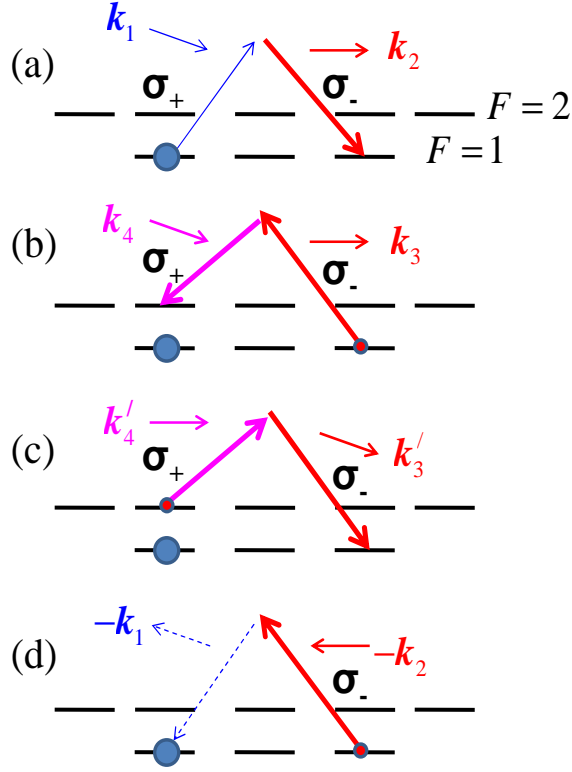}%
\caption{(Color online) Example of a possible excitation scheme. The Raman
transitions connect the Zeeman sublevels of $F=1$, $F=2$ hyperfine levels.
Each Raman transition involves a $\sigma_{+}$ and a $\sigma_{-}$ polarized
beam. }%
\label{fig3a}%
\end{center}
\end{figure}

A critical issue in a multilevel system is the ability to selectively drive
the target transitions. Field polarization can be used to provide the
selectivity if the angle between the fields is small (or close to $\pi$); in
this limit, one can use circularly polarized fields, as shown in Fig.
\ref{fig3a}. The ground state manifold consists of $F=1$ and $F=2$ hyperfine
states and the $\Lambda$-systems of the Raman protocol involve magnetic
substates of these levels. Since the angle between the beams is assumed to be
on the order 10 mrad or so, the fields can taken to be cross-polarized, in
first approximation.

Optical pumping by the strong control field determines the uncoupled initial
state ($F=1$, $m=-1$). In step (a), fields 1 and 2 create a Raman coherence
$\rho_{1,-1:1,1}$, where the notation is $\rho_{F,m:F^{\prime},m^{\prime}}$.
The Doppler dephasing is then stopped by a $\pi$-pulse, composed of two beams
propagating along $\mathbf{k}_{3}$ and $\mathbf{k}_{4}$ in such a way that
$\mathbf{k}_{4}-\mathbf{k}_{3}=\mathbf{k}_{1}-\mathbf{k}_{2}$ [step (b)].
These fields convert the coherence $\rho_{1,-1:1,1}$ into $\rho_{1,-1:2,-1}$.
The process in which fields 4 and 3 drive Raman transitions between states
$\left\vert 1,-1\right\rangle $ and $\left\vert 1,1\right\rangle $ is
suppressed, provided the hyperfine frequency separation is much larger the
inverse pulse duration. In step (c) a Raman $\pi$ pulse having $\mathbf{k}%
_{4}^{\prime}-\mathbf{k}_{3}^{\prime}=-\left(  \mathbf{k}_{4}-\mathbf{k}%
_{3}\right)  $ restores the $\rho_{1,-1:1,1}$ coherence and prepares the
system in such a fashion that the subsequent application of a control pulse
having propagation vector $-\mathbf{k}_{2}$ restores the \ input signal field
propagating in the $-\mathbf{k}_{1}$ direction at time $t=t_{3}+t_{12}$.

\section{Conclusion}

We have proposed a Raman based quantum memory scenario that operates with
close to 100\% efficiency. This protocol fits well to intermediate time
scales, with signal duration of order 100ns, a time range that is well adapted
to experiments based on high spectral purity continuous wave laser sources. We
have shown that the spectral components are stored in different atomic
velocity classes and have explained how to optimize the optical depth
depending of the needed storage bandwidth. Since we have assumed that the
signal field is weak and since spontaneous emission is negligible in this
protocol, we expect the results to be unchanged if the classical input field
is replaced by a quantized, pulsed radiation field. 

\acknowledgments
We are pleased to acknowledge stimulating discussions with E. Giacobino, G. Leuchs, J. Laurat and T. Chaneli\`ere. PRB would like to thank l'Institut Francilien de recherche sur les atomes froids (IFRAF) for helping to support his visit to Laboratoire Aim\'{e} Cotton and for the hospitality shown to him during his visit.


\begin{thebibliography}{99}
\bibitem {pop}In the MK protocol one prevents the control field from
interacting with the populated level, not only to ensure uniform illumination
of the medium, but also to avoid excitation to the upper level that would
entail amplification and spontaneous emission noise. This cannot occur in the
present protocol.
\bibitem {doppler}This connection between the inhomogeneous phase shift and
spatial position is unique to the Doppler effect. The phase shift is connected
simply to the distance travelled by the atoms. As a consequence, even a short
laser pulse can result in an accumulated Doppler phase shift for the atoms.
This rather paradoxical feature does not occur when the inhomogeneous
broadening is caused by position dependent energy level shifts of stationary
atoms, as in a solid.
\end{thebibliography}
\end{document}